\documentclass[]{iopart}
\usepackage{iopams} 

\begin{document}
\title[]
{An interacting gauge field theoretic model for the Hodge theory: basic  canonical brackets}

\author{R. Kumar$^1$\footnote{Present address: S. N. Bose National Centre for Basic Sciences, 
Kolkata$-$700 098, India}, S. Gupta$^1$\footnote{Present address: The Institute of Mathematical Sciences, 
Chennai$-$600 113, India}, R. P. Malik$^{1, 2}$}

\address{$^1$Department of Physics, Centre of Advanced Studies,
           Banaras Hindu University, Varanasi$-$ 221 005, (U. P.), India}
           
\address{$^2$DST Centre for Interdisciplinary Mathematical Sciences,
           Faculty of Science, Banaras Hindu University, Varanasi$-$ 221 005, (U. P.), India }   
      
$~~~~~~~~~~~~~~~~${\footnotesize{E-mails: raviphynuc@gmail.com; saurabh@imsc.res.in; rpmalik1995@gmail.com}}

\begin{abstract}
We derive the basic canonical brackets amongst the creation and annihilation
operators for a two $(1 + 1)$-dimensional (2D) gauge field theoretic model of an
 interacting Hodge  theory  where a $U(1)$ gauge field ($A_\mu$) is coupled with the 
fermionic Dirac fields ($\psi$ and $\bar \psi$). In this derivation, we exploit the spin-statistics 
theorem, normal ordering and the strength of the underlying {\it six} infinitesimal continuous symmetries 
(and the concept of their generators) that are present in the theory. We do not use the 
definition of the canonical conjugate momenta (corresponding to the basic fields of the theory) 
{\it anywhere} in our whole discussion. Thus, we conjecture that our present approach provides 
an alternative to the canonical method of quantization for a class of gauge field 
 theories that are physical examples of Hodge theory where the continuous 
symmetries (and corresponding generators) provide the physical realizations of the de Rham 
cohomological operators of differential geometry at the algebraic level.
\end{abstract}

\vspace{2pc}
 

\maketitle

\section{Introduction}

One of the earliest methods of quantization scheme (for a given physical system) 
is the canonical method of quantization where a set of {\it three} basic ideas 
is primarily exploited {\it together}. First of all, by using the spin-statistics theorem, we 
differentiate between the commuting (bosonic) and anticommuting (fermionic) 
variables of a given Lagrangian. Thereafter, we define the canonical conjugate 
momenta corresponding to the basic dynamical variables of the theory. We obtain 
the basic  (graded) Poisson brackets amongst the dynamical variables and corresponding 
conjugate momenta at the {\it classical} level. These are, finally, promoted to 
the basic (anti)commutators at the {\it quantum} level and the quantization of the system ensues. 
If the physical system supports the existence of creation and annihilation operators, then, 
the normal ordering is required for the physical quantities to make some {\it useful} sense.

In the field theoretic models, we define a Lorentz scalar Lagrangian density for a given 
physical system. We follow the above sequence for the procedure of quantization. 
That is to say, we distinguish between the bosonic and fermionic fields 
(by spin-statistics theorem) and define the canonical conjugate momenta corresponding 
to the basic dynamical fields. The basic (graded) Poisson brackets of the theory at  {\it classical} level 
are promoted to the canonical (anti)commutators at the {\it quantum} level. 
However, in the field theoretic models, the fields (and corresponding momenta) 
are operators which are expressed, normally, in terms of the creation and annihilation 
operators. The above cited canonical {\it quantum} (anti)commutators are, at 
this stage, usually  expressed in terms of the creation and annihilation operators. In other words, these
basic canonical  brackets are written in the language of the above creation and annihilation operators. 
Physical quantities of interest (e.g. Hamiltonian, conserved charges, etc.) 
are expressed in terms of the creation and annihilation operators, too. However, 
to avoid the unwanted infinities, the operators (present in the above physical quantities)
are normal ordered so that  they could make some  physical sense [1,2].

In our present investigation, we shall utilize the virtues of normal ordering 
and spin-statistic theorem. However, we shall {\it not} take the help of {\it mathematical} 
definition of the canonical conjugate momenta for the basic dynamical fields of 
our theory (i.e. 2D interacting $U(1)$ gauge theory of photon and Dirac fields). 
Rather, in the place of the latter, we shall utilize the beauty and strength of the
{\it physical} symmetry principles (and the concept of a generator for a given 
infinitesimal continuous symmetry).         
Our present method of quantization, eventhough algebraically more involved, 
is physically more appealing  because it is the symmetry properties of our theory 
that play a key role in our computations of the basic canonical brackets. 
In contrast,  it is the {\it mathematical} definition of the canonical momentum 
that plays a decisive role in the determination  of the basic brackets in the 
case of canonical method of quantization.

Recently we have exploited the central theme of our approach to quantize the 2D free Abelian 
1-form and 4D free Abelian 2-form gauge theories where the {\it mathematical} 
definition of canonical conjugate momenta has {\it not} been used anywhere [3]. It is 
the symmetry properties of the above field theoretic models for the Hodge theory  [4--7]
(and their corresponding generators) that have played a decisive role in the derivation 
of the basic canonical brackets amongst the creation and annihilation operators of 
the above theories which contain bosonic as well as fermionic operators. Our present model 
of 2D quantum electrodynamics (QED) is a field theoretic example of an {\it interacting} 
Hodge theory (see, e.g. [8,9]) because the symmetries provide the physical realizations 
of de Rham cohomological operators [10--12]. It is a challenging endeavor to check the sanctity of our method of 
quantization [3] in the context of our present {\it interacting} theory as well. Of course, the interacting theory
is more general than its free counterpart because the latter is a limiting case of the former.

Our present endeavour is essential on the following grounds. First and foremost, it is very 
urgent problem for us to extend our method of quantization (that is valid for the 2D Abelian 
1-form and 4D Abelian 2-form {\it free} theories) to an {\it interacting} model for the Hodge
 theory where there is an interaction between the matter fields and gauge field. Second, 
it is always important and interesting to provide an alternative to the {\it mathematical} 
definition in the language of some basic physical properties. For instance, we provide an 
alternative to the mathematical definition of the canonical conjugate momenta in terms of the continuous 
symmetries and concept for the generator of a continuous symmetry 
transformation. Finally, our present endeavor adds a new dimension to the utility 
of symmetry principles (for a class of gauge field theories that turn out to be the field  
theoretic  models for the Hodge theory) because it encompasses in its ever--widening  folds
the basic canonical brackets, too.

The contents of our present investigation are organized as follows. First of all, 
we discuss a set of {\it six} continuous symmetries of a 2D QED which is dynamically 
closed system of a $U(1)$ gauge field and Dirac fields in Sec. 2. Our forthcoming Sec. 3 
contains the derivation of canonical basic brackets from the (anti-)BRST symmetries and 
corresponding generators. We derive the same basic brackets by exploiting the (anti-)co-BRST 
symmetries in Sec. 4. Our Sec. 5 deals with a concise derivation of the above brackets by 
exploiting the unique  bosonic symmetry of the theory. For the sake of comparison, we discuss, 
in a concise manner, the canonical method of quantization for our present theory in 
Sec. 6. Finally, we make some concluding remarks and point out a few
future directions for further investigations in our Sec. 7.

In our Appendix A, we derive the basic brackets amongst the creation and annihilation operators 
that appear in the normal mode expansions of the Dirac fields in their full generality (where we do not  
take the help of the condition $t = 0$).  
We also establish that the canonical brackets at the field level are equivalent to 
the canonical brackets at the level of the creation and annihilation operators.

\section{Preliminaries: Lagrangian formalism} 
In this section, we discuss various continuous symmetries of the 2D QED where there is an interaction 
between the $U(1)$ gauge field ($A_\mu$) and the Dirac fields ($\psi$ and $\bar \psi$). 
We begin with the following locally gauge invariant Lagrangian density (see, e.g. [1,2])
\begin{eqnarray}
\fl \qquad \qquad {\cal L} _0 = - \frac{1}{4} \; F^{\mu\nu}\, F_{\mu\nu} 
+ \bar\psi \, (i \gamma^\mu \,D_\mu - m ) \, \psi, 
\end{eqnarray}
where $F_{\mu\nu} = \partial_\mu A_\nu - \partial_\nu A_\mu$ is derived from the 2-form $F^{(2)} = d\,A^{(1)}$.
In the above, $d = dx^\mu\, \partial_\mu$ (with $d^2 = 0$) is the exterior derivative and the 1-form 
$A^{(1)} = dx^\mu\, A_\mu$ defines the vector potential ($A_\mu$). We have taken 
$D_\mu \,\psi = \partial_\mu\, \psi + i\, e\, A_\mu\, \psi$ as the covariant derivative on the Dirac field $\psi$.
The above Lagrangian density is a closed system of a massless vector gauge boson and Dirac fields where 
the interaction term is ($- e\, \bar \psi\, \gamma^\mu\, A_\mu\, \psi$). This derivation 
is true in any arbitrary dimension of spacetime. Here the quantity $e$ is the electric charge of the Dirac fields.

In the specific two $(1+1)$-dimensions of spacetime, $F_{\mu\nu}$ has only one non-vanishing 
component which is nothing but the electric field $E = - \varepsilon^{\mu\nu}\, \partial_\mu A_\nu 
= \partial_0 A_1 - \partial_1 A_0$ where we have taken the Levi-Civita tensor $\varepsilon_{\mu\nu}$ 
with the conventions $\varepsilon_{01} = +1 = - \varepsilon^{01}$, 
$\varepsilon_{\mu\nu}\,\varepsilon^{\nu\lambda} = \delta^\lambda_\mu,$ etc., and 
the Greek indices $\mu,\,\nu, \lambda,... = 0, 1$.
One of most elegant approaches to quantize the above gauge theory is the Becchi--Rouet--Stora--Tyutin (BRST)
formalism where the gauge--fixing and Faddeev--Popov ghost terms are incorporated in the Lagrangian density. 
Such an (anti-)BRST invariant 2D Lagrangian density, in the Feynman gauge, is as follows (see, e.g. [8,9] for details) 
\begin{eqnarray}
{\cal L} _b &=& {\cal L} _0  - \frac{1}{2} \; (\partial \cdot A)^2 
- i \, \partial_\mu \bar C \; \partial^\mu C \nonumber\\
&\equiv &  \frac{1}{2} \; E^2 + \bar\psi \, (i \,\gamma^\mu \,D_\mu - m ) \, \psi  
- \frac{1}{2} \; (\partial \cdot A)^2 - i \, \partial_\mu \bar C \, \partial^\mu C,
\end{eqnarray}
where $(\bar C)C$ are the fermionic ($\bar C^2 = C^2 = 0, \; C \, \bar C + \bar C \, C = 0$) (anti-)ghost
fields, the Dirac fields ($\psi, \bar \psi$) obey the anticommutativity property 
($ \psi\, \bar \psi + \bar \psi\, \psi = 0)$ at the {\it classical} level and  
$ \bar \psi = \psi^\dagger \gamma^0$. Here $\gamma^\mu = (\gamma^0, \, \gamma^1)$ are the $2 \times 2$ Dirac gamma matrices in 2D and one can choose them in terms of the Pauli $\sigma$-matrices as $\gamma^0 = \sigma_2$,
$\gamma^1 = i\, \sigma_1$ so that $\gamma_5 = \gamma^0\, \gamma^1 = \sigma_3$. It can be readily checked that 
$\{\gamma^\mu,\; \gamma^\nu\} = 2 \eta^{\mu\nu}$ and $\gamma_\mu \, \gamma_5 = \varepsilon_{\mu\nu}\,\gamma^\nu$
where $\eta_{\mu\nu} =$ diag $(+1,\, -1)$ is the flat 2D Minkowski metric.
 We adopt  here the convention of left derivative w.r.t. the fermionic fields 
($\psi, \, \bar \psi, \, C, \, \bar C$) so that we obtain the conjugate momenta as:
$\Pi_\psi = - \; i \; \psi^\dagger,\; \Pi_C = i \; \dot {\bar C}, \; \Pi_{\bar C} = - \;i \; \dot C$.

The Lagrangian density (2) respects the following on-shell ($\Box C = \Box \bar C = 0$) nilpotent
($s^2_{(a)b} =0$) (anti-)BRST symmetry transformations ($s_{(a)b}$) (see, e.g. [8,9])
\begin{eqnarray}
\fl \qquad && s_{ab} A_\mu = \partial_\mu \bar C, \;\;\quad s_{ab} \bar C = 0, \;\;\quad 
s_{ab} C = +\, i\, \big(\partial\cdot A\big),\;\;\quad s_{ab} E = 0,\nonumber\\
\fl  \qquad  && s_{ab} \psi = - i \; e \; \bar C \psi, 
\qquad s_{ab} \bar \psi = i \; e \; \bar C \bar\psi \equiv - i \, e \, \bar\psi\,\bar C,
  \nonumber\\
\fl \qquad && s_b A_\mu = \partial_\mu C, \qquad s_b C = 0, \qquad s_b \bar C = -\, i \,\big(\partial \cdot A\big),
\qquad s_b E = 0, \nonumber\\
\fl \qquad &&s_b \psi = - i \; e \; C \psi,  
\qquad s_b \bar \psi = i \; e \; C \bar \psi \equiv - i \, e \, \bar \psi\, C,
\end{eqnarray}
where we have used, at a couple of  places, the anticommutativity  property 
($C\,\psi + \psi \, C = 0,\; \bar C\, \bar \psi + \bar \psi \, \bar C = 0,$ etc.) of the 
fermionic ($\psi^2 = {\bar \psi}^2 = C^2 = \bar C^2 = 0$) fields ($\psi, \, \bar \psi, \, C, \, \bar C$). 
It is straightforward to check that the above on-shell nilpotent  (anti-) BRST symmetry transformations are absolutely 
anticommuting ($s_b\, s_{sb} + s_{ab}\, s_b = 0$) on the on-shell where $\Box C = \Box \bar C = 0$. 
The corresponding on-shell nilpotent and conserved ($\dot Q_{(a)b} = 0$) (anti-)BRST charges, 
that generate the symmetry transformations (3), are
\begin{eqnarray}
Q_{ab} = \int dx \; [ E \, \partial_1 \bar C - e \, \bar\psi\, \gamma_0 \,  \bar C\, \psi
- (\partial \cdot A) \, \dot {\bar C}],\nonumber\\
Q_b = \int dx \; [ E \, \partial_1 C - e \, \bar\psi\, \gamma_0\,  C \,\psi - (\partial \cdot A) 
\, \dot C], 
\end{eqnarray}
which can be re-expressed in a simpler form if we use the equation of motion 
$\partial_\mu F^{\mu\nu} + \partial^\nu (\partial \cdot A) = e \; \bar \psi \gamma^\nu \psi$. In other words,
we {\it also} have the simpler forms: $Q_{ab} = \int dx\;[\partial_0 (\partial \cdot A)\, \bar C 
- (\partial \cdot A)\, \dot {\bar C}]$ and $Q_b = \int dx\;[\partial_0 (\partial \cdot A)\, C 
- (\partial \cdot A)\, \dot C].$

We note that under the following on-shell ($\Box C = \Box \bar C = 0$) nilpotent ($s^2_{(a)d} = 0$)
(anti-)co-BRST symmetry transformations ($s_{(a)d}$) (see, e.g. [8,9])
\begin{eqnarray}
\fl \qquad && s_{ad} A_\mu = -\varepsilon_{\mu\nu}\, \partial^\nu  C, \qquad s_{ad} C = 0, \qquad 
s_{ad} \bar C = i\, E, \qquad s_{ad}\big(\partial \cdot A\big) = 0, \nonumber\\
\fl \qquad && s_{ad}  \psi  = - \, i \, e \, C \, \gamma_5 \, \psi, \qquad
s_{ad}  \bar \psi = i \, e \, \bar \psi\, C \, \gamma_5, \nonumber\\
\fl \qquad && s_d A_\mu = -\varepsilon_{\mu\nu}\, \partial^\nu \bar C, \qquad s_d \bar C = 0, \qquad 
s_d C = - \,i\, E, \qquad s_d \big(\partial \cdot A\big) = 0, \nonumber\\
\fl \qquad && s_d \psi  = - \, i \, e \, \bar C \, \gamma_5 \, \psi, \qquad 
s_d \bar \psi =  i \, e \, \bar \psi\,\bar C \, \gamma_5,
\end{eqnarray}
the Lagrangian density (2) transforms to a total spacetime derivative thereby showing the 
symmetry property of the action integral ($S = \int d^2x\; {\cal L}_b$) due to the Gauss's 
divergence theorem. It should be noted that the transformations (5) are {\it symmetry} 
transformations for the Lagrangian density (2) only in the massless limit ($m = 0$). 
In other words, the above symmetry (5) is true for the {\it chiral} fermions only.
According to Noether's theorem, we have conserved charges $Q_{(a)d}$ 
corresponding to the above symmetry transformations (5) as:
\begin{eqnarray}
\fl \qquad Q_{ad} &=& \int dx \;\Big[E \; \dot C + e \; \bar \psi \; \gamma_1 \; C \; \psi 
+ \partial_1 (\partial \cdot A) \; C \Big ],\nonumber\\
\fl \qquad Q_d &=& \int dx \; \Big[E \; \dot {\bar C} + e \; \bar \psi \; \gamma_1 \; \bar C \psi 
+ \partial_1 (\partial \cdot A) \; \bar C \Big ],
\end{eqnarray}
which turn out to be the generators of the transformations (5). These charges can be written
in a simpler form if we use the equation of motion $\partial_\mu F^{\mu\nu} 
+ \partial^\nu (\partial \cdot A) = e \; \bar \psi \gamma^\nu \psi$
to re-express (6) as  $Q_{ad} = \int dx\;[E\, \dot C - \dot E \, C]$ and 
$Q_d = \int dx\;[E\, \dot {\bar C} - \dot E \, \bar C]$.

There is a unique (i.e. $\{s_b,\; s_d\} = s_\omega = - \{s_{ab},\; s_{ad}\}$) bosonic symmetry $(s_\omega)$
in our theory which is obtained from the anticommutators of $s_{(a)b}$ and $s_{(a)d}$ [8,9,4]. Under this symmetry transformation, the relevant fields of the theory transform as:
\begin{eqnarray}
\fl \qquad s_\omega A_\mu = \partial_\mu E- \varepsilon_{\mu\nu}\,\partial^\nu (\partial \cdot A),
\qquad s_\omega \psi = - \, i \, e \big [\gamma_5 (\partial \cdot A) + E \big ]  \psi, \nonumber\\
\fl \qquad s_\omega C = 0, \qquad s_\omega \bar C = 0, 
\qquad s_\omega(\partial \cdot A) = \Box E, \qquad  s_\omega E = \Box (\partial \cdot A), \nonumber\\
\fl \qquad s_\omega \bar \psi = + \, i \, e \big [\gamma_5\, (\partial \cdot A) + E \big ] \bar \psi
\equiv i\,e\, \bar \psi \big[E - \gamma_5 \, (\partial \cdot A)\big].
\end{eqnarray} 
It can be checked that the Lagrangian density (2) transforms to a total spacetime derivative under (7).
As a consequence, the action integral remains invariant under the infinitesimal and continuous symmetry  
transformations (7). The conserved charge corresponding to the above transformations (7) is
\begin{eqnarray}
\fl \qquad && Q_\omega = \int dx \; \Big [(\partial \cdot A) \partial_1 (\partial \cdot A) 
- E \; \partial_1 E - e \; (\partial \cdot A) \bar \psi \; \gamma_1 \; \psi 
+ e \; E \; \bar \psi \; \gamma_0 \; \psi \Big ], 
\end{eqnarray} 
which turns out to be the generator of transformations (7).

Finally, we have a  global ghost-scale symmetry in the theory where $C \to e^{+\Lambda} \,C$, 
$\bar C \to e^{-\Lambda} \,\bar C$. Here $\Lambda$ is an infinitesimal 
spacetime independent scale parameter. The infinitesimal version of this transformation ($s_g$) is   
\begin{eqnarray}
\fl \qquad s_g C = + C, \qquad s_g \bar C = - \bar C, \qquad s_g\Phi = 0, \qquad \Phi = A_\mu, \, \psi,\, \bar \psi, 
\end{eqnarray} 
where, for the sake of brevity, we have set $\Lambda = 1$. The corresponding 
Noether's conserved charge ($Q_g$) is as follows
\begin{eqnarray}
\fl \qquad Q_g  = i\,\int dx\;\big[C \, \dot {\bar C} + \bar C \, \dot C \big]. 
\end{eqnarray} 
The above charge $Q_g$ is the generator of the transformations (9). Thus, as claimed   earlier, we have {\it six}
infinitesimal and continuous symmetries in the theory.

The decisive features of the on-shell nilpotent (anti-)BRST and (anti-)co-BRST symmetries are 
the invariances of the kinetic and gauge--fixing terms of the Lagrangian density (2), respectively. 
The ghost term, on the other hand,  remains invariant under the bosonic symmetry transformations. 
The key feature of the ghost-scale symmetry is the observation that only the (anti-)ghost 
fields transform globally   and  rest of the fields of the theory remain unchanged under it. 
These symmetries, at the algebraic level, provide the physical realizations of the de
Rham cohomological operators of differential geometry (see, e.g. [8,9,4] for details). As a consequence,
our present 2D interacting theory (i.e. 2D QED) is a physical model for the Hodge theory.

\section{(Anti-)BRST symmetries: basic brackets}
In our earlier work on the 2D Abelian 1-form gauge theory [3], we have exploited the ideas of 
symmetry principles (along with spin--statistics theorem and normal ordering) 
to derive the basic non--vanishing canonical brackets amongst the creation and 
annihilation operators of the gauge field $A_\mu$ and (anti-)ghost fields $(\bar C)C$ as:
\begin{eqnarray}
&& \big [a_\mu (k),\; a^\dagger_\nu (k') \big ] = \eta_{\mu \nu} \, \delta (k - k'), \quad
\big\{ c (k), \;\bar c^\dagger (k') \big\} = +\, i \, \delta (k - k'), \nonumber\\
&& \big\{ c^\dagger (k), \;\bar c (k') \big\} = - \,i \, \delta (k - k'),  
\end{eqnarray}
where the above operators are present in the normal mode expansions of the basic fields of the
Lagrangian density (2) (in the limit $\psi =0, \; \bar \psi = 0$) as (see, e.g. [1,3]): 
\begin{eqnarray}
A_\mu (x,t) &=& \int \frac {dk}{\sqrt {2 \pi \,2 k_0}} \; \Big [a_\mu (k) \, e^{+ i k \cdot x} 
 +  a^\dagger_\mu (k) \, e^{- i k \cdot x} \Big ], \nonumber\\
C(x,t) &=& \int \frac {dk}{\sqrt {2 \pi\, 2 k_0}} \; \Big [c(k) \, e^{+ i k \cdot x} 
+  c^\dagger (k) \, e^{- i k \cdot x} \Big ], \nonumber\\  
\bar C(x,t) &=& \int \frac {dk}{\sqrt {2 \pi\, 2 k_0}} \; \Big [\bar c(k) \, e^{+ i k \cdot x} 
 +   \bar c^\dagger (k) \, e^{- i k \cdot x} \Big ].
\end{eqnarray}
Here the two--vector $k_\mu = (k_0,\; k_1 = k)$ is the momentum vector and 
the dagger operators $ a^\dagger_\mu (k), c^\dagger (k)$ and $\bar c^\dagger (k)$ are the creation operators
for a photon, a ghost and an anti-ghost quanta, respectively. The non-dagger operators 
$ a_\mu (k), c(k)$ and $\bar c(k)$ stand for the corresponding annihilation operators
for a single quantum. It is also clear that the operators ($a^\dagger_\mu (k), a_\mu (k)$)
are {\it bosonic} in nature as against the operators ($ c(k), \bar c(k), c^\dagger (k), \bar c^\dagger (k)$)
that are {\it fermionic}. The (anti-)BRST symmetries (being supersymmetric type in nature), 
our present {\it interacting} theory is endowed with bosonic as well as fermionic creation 
and annihilation operators.

In the above derivation, we have mainly utilized the definition of the generator 
of a continuous symmetry transformation. According to the common folklore in quantum 
field theory, the conserved charges
(that are derived due to the presence of the continuous symmetries in the theory)
generate the infinitesimal and continuous symmetry transformations as 
\begin{eqnarray}
s_r \Phi = \pm \;i \; \big [\Phi, \; Q_r \big ]_\pm,\;
\qquad r \; = \; b, \;  ab, \;  d, \; ad, \; \omega, \; g,
\end{eqnarray}
where $\Phi$ is the generic field of the theory and $Q_r$ are the conserved charges. The $(\pm)$ signs, as the subscripts on the square bracket,
correspond to the (anti)commutator for the generic field $\Phi$ being (fermionic)
bosonic in nature. The  $(\pm)$ signs, in front of the expression on the r.h.s.
(i.e. $\pm \; i \;\big [\Phi, \; Q_r \big ]_\pm $), need explanation. 
The pertinent points, regarding the choice of a specific sign for
a specific purpose, are as follows:
\begin{enumerate}
\item for $s_r = s_b, \; s_{ab}, \; s_d, \; s_{ad}$ only the negative sign is
to be taken into account (i.e. $s_b A_\mu = - i \; \big [A_\mu, \; Q_b \big ],
 \; s_b \bar C  = - i \; \{ \bar C, \; Q_b \} $, etc.), and
  
\item for $s_r = s_g, \; s_\omega $ the negative sign is to be taken into account 
for the bosonic field and the positive sign is to be chosen for the fermionic 
field  (e.g. $s_g A_\mu = - i\; \big [ A_\mu, \; Q_g \big ], 
s_g C = + i \; \big [ C, \; Q_g \big ] , \;
s_g {\bar C} = + i \; \big [\bar C, \; Q_g \big ],$ etc.).
\end{enumerate}
In the derivation of the non--vanishing basic brackets (11), we have utilized the above 
rules in the computation of the basic quantum (anti)commutators. It is obvious that the
rest of brackets are zero for the free theory [as far as the non--vanishing brackets 
(11) and related brackets are concerned].

In our present endeavor, we shall focus on the derivation of the basic brackets for 
the fields $\psi$ and $\bar \psi$. Following (3) and (13), it is clear that the 
Dirac fields transform, under the on-shell nilpotent (anti-)BRST symmetries,  as 
\begin{eqnarray}
\fl \qquad && s_b \psi = - i \; \{\psi, \;  Q_b\} = - i \; e \;  C \psi, \qquad
s_{ab} \psi = - i \; \{\psi, \; Q_{ab}\} = - i \; e \; \bar C \psi, \nonumber\\
\fl \qquad && s_b \bar\psi = - i \; \{\bar\psi, \; Q_b\} = + i \; e \; C \bar\psi, \qquad s_{ab} \bar\psi 
= - i \; \{\bar\psi, \; Q_{ab}\} = + i \; e \; \bar C \bar\psi,
\end{eqnarray}
where we have used only the basic concepts of the continuous symmetries (and their 
generators) as well as the spin-statistics theorem.
We can, at this stage, take the following mode expansions for the Dirac fields in the momentum 
phase space [13]
\begin{eqnarray}
\fl \quad \psi (x, t) =  \int \frac{ d k }{\surd (2\pi \, 2k_0)} \sum_{\alpha} 
\Bigl( b^\alpha (k) \,  u^\alpha(k) \; e^{+ i k \cdot x} 
+  (d^\alpha)^\dagger (k) \, v^\alpha(k) \, e^{- i k \cdot x} \Bigr),
\end{eqnarray}
\begin{eqnarray}
\fl \quad \psi^\dagger (x, t) =  \int \frac{ d k }{\surd (2\pi \; 2k_0)} \sum_{\alpha} 
\Bigl( (b^\alpha)^\dagger (k) \; (u^\alpha)^\dagger (k) \, e^{- i k \cdot x}
+ d^\alpha (k) \; (v^\alpha)^\dagger (k) \, e^{+ i k \cdot x} \Bigr),
\end{eqnarray}  
where $(b^\alpha)^\dagger (k) , (d^\alpha)^\dagger (k)$ are the creation operators 
and the corresponding operators without dagger (i.e. $b^\alpha(k), d^\alpha(k)$) are 
the annihilation operators. All these operators are {\it fermionic} in nature. The bosonic 
variables $u^\alpha (k)$ and $v^\alpha (k)$ are the plane wave solutions of the Dirac 
equation for the positive and negative energies, respectively. We shall be exploiting some of the 
key properties associated with these operators. A few of these 
(that are useful to our current endeavor) are  listed below (see, e.g. [13])
\begin{eqnarray}
\fl \qquad && \sum_\alpha u^\alpha (k) \; \bar u^\alpha (k) = ( \gamma^\mu k_\mu + m ), \qquad
\sum_\alpha v^\alpha (k) \; \bar v^\alpha (k) = ( \gamma^\mu k_\mu - m ),\nonumber\\ 
\fl \qquad && (u^\beta)^\dagger (k) \; u^\alpha (k)  =  2 k_0 \; \delta^{\alpha \beta}, \hskip 2cm
 (v^\beta)^\dagger (k) \; v^\alpha (k)  = 2 k_0 \; \delta^{\alpha \beta},
\end{eqnarray}
where $\bar u^\alpha (k) = (u^\alpha)^\dagger (k) \; \gamma^0, \quad 
\bar v^\alpha (k) = (v^\alpha)^\dagger (k) \; \gamma^0 $.

It will be worthwhile to mention that, in the case of 2D {\it free} U(1) gauge theory 
(i.e. $\psi = \bar \psi = 0$), it was 
quite easy to express all the {\it conserved charges} in terms of the creation and annihilation 
operators in a compact form without any exponentials  because, in these 
expressions, the field variables turned out to be {\it quadratic} (bilinear) only (see, e.g. [3] for details). 
This is, however, not the case as far as the interacting 2D QED is concerned. 
It can be seen that, in the expressions for $Q_b$ and $Q_{ab}$ [cf. (4)], we have terms 
(e.g. $\int dx \; e C \psi^\dagger \psi $ etc.), that are {\it not} quadratic. 
Thus, to obtain the canonical brackets between the fermionic annihilation and 
creation operators (e.g. $b^\alpha (k), (b^\alpha)^\dagger (k)$, etc.),
we have to adopt a different technique because non-quadratic terms can not be expressed in
a compact form in terms of the creation and annihilation operators without any exponentials.
To elucidate this point, we do the following explicit exercise which conveys the main 
ideology of our approach.

Let us try to obtain the canonical anticommutators from the following principle of the 
continuous symmetry transformation [cf. (13)]: 
\begin{eqnarray}
s_b \; \psi \; = \; - \; i \; e \;  C \; \psi \; = \; - \; i \; \bigl \{ \psi,\; Q_b \bigr \} \; 
\Longrightarrow \; e \; C \; \psi \; = \; \bigl \{ \psi, \; Q_b \bigr \}, 
\end{eqnarray}
where the BRST charge $Q_b$ is the generator of the transformation $s_b$. Since 
the nilpotent (anti-)BRST charges are conserved quantities, it is simpler to perform
all the computations with $t = 0$ (see, e.g. [13] for details). 
The results, thus obtained, would be same as the ones obtained in all 
their generality [i.e. without taking the limit $t = 0$ (see, e.g., Appendix A below)]. Exploiting 
the mode expansions of the basic fields given in (12) and (15), in this limit (i.e. $t = 0$), the l.h.s. is 
\begin{eqnarray}
\fl \qquad && e \int \frac { d k \; d k^\prime}{ (2 \pi ) \surd (2 k_0 \; 2 k^\prime_0 )}
\sum_\alpha \Bigl( c(k) \; b^\alpha (k^\prime) \; u^\alpha (k^\prime)\; 
exp \;[- i \; (k + k^\prime ) \; x]\nonumber\\ 
\fl \qquad && + c(k) \; (d^\alpha)^\dagger (k^\prime) \; v^\alpha (k^\prime) \; 
exp \;[- i \; (k - k^\prime ) \; 
 x] + c^\dagger (k) \; b^\alpha (k^\prime)\;  u^\alpha (k^\prime) \nonumber\\
\fl \qquad && exp \;[ + i \; (k - k^\prime ) \; x] + c^\dagger (k) \; (d^\alpha)^\dagger (k^\prime) 
\; v^\alpha (k^\prime) \; exp \;[ + i \; (k + k^\prime ) \; x] \Bigr),
\end{eqnarray}
where, in the exponentials, we have only the space part of the 2D dot products 
and $k$ and $k^\prime$ correspond to the space part of the momenta for the fields $C(x)$ and 
$\psi (x)$ in the phase space. The powers of exponential play very important role when we 
compare the l.h.s. with r.h.s. (while exploiting the basic equation (18) related to the
key tenets of continuous symmetries and their generators).

As far as the computation of the r.h.s. is concerned, it should be noted that the following 
portions of the BRST and anti-BRST charges, namely;
\begin{eqnarray}
\fl \qquad Q^{(R)}_b = e \int dx \; C(x) \; \psi^\dagger (x) \; \psi (x), \qquad 
Q^{(R)}_{ab} =  e \int dx \; \bar C(x) \; \psi^\dagger (x) \;  \psi (x),
\end{eqnarray}
contribute in the computations of $s_{(a)b} \psi$ and $s_{(a)b} \bar \psi$. 
Here the superscript $(R)$ on the charges (i.e. $Q^{(R)}_{(a)b}$) denotes  the relevant portion 
of the conserved and nilpotent charges $Q_{(a)b}$ [cf. (4)]. Furthermore, 
it should be emphasized that the rest of the quadratic parts of the (anti-)BRST 
charges ($Q_{(a)b}$) can be expressed in terms of the creation and annihilation 
operators and they appear in a compact form as is the case in [3]. The explicit form 
of the contributing factor (in the case of $s_b \psi$), from the r.h.s., is 
\begin{eqnarray}
e \int  dy \; \{\psi(x),\;  C(y) \; \psi^\dagger(y) \; \psi(y)\}.
\end{eqnarray}
The above expression can be written in terms of the mode expansions of the fields 
$ C(y), \psi (y), \psi^\dagger (y)$ and $\psi (x)$ by exploiting (12), (15) and (16). It 
is evident that the comparison of the l.h.s. with the r.h.s. leads to the cancellation of 
the factor  $``e"$ present on both the sides [cf. (19), (21)].

The relevant $Q_b$ part (i.e. $\int dy \;C(y) \; \psi^\dagger(y) \; \psi(y)$) for 
our computation, in terms of the mode expansion, can be written (for $ t = 0$) as
\begin{eqnarray}
\fl \qquad && \sum_\gamma \sum_\sigma \int dy \int \frac{dk_2 \; dk_3 \; dk_4}{(2\pi )^{3/2} 
\surd {(2k^0_2 \; 2k^0_3 \; 2k^0_4)}} \Bigl( a_1 \;
exp \;[- i \; (k_2 - k_3 + k_4) \; y] \nonumber\\
\fl \qquad && + \; a_2 \;exp\;[- i \; (k_2 - k_3 - k_4) \; y] + \; a_3 \;
exp \;[- i \; (k_2 + k_3 + k_4) \; y] \nonumber\\
\fl \qquad && + \; a_4 \;exp\;[- i \; (k_2 + k_3 - k_4) \; y] + \; a_5 \;
exp \;[+ i \; (k_2 + k_3 - k_4) \; y] \nonumber\\
\fl \qquad && + \; a_6 \;exp\;[+ i \; (k_2 + k_3 + k_4) \; y] + \; a_7 \;
exp\;[+ i \; (k_2 - k_3 - k_4) \; y] \nonumber\\
\fl \qquad && + \; a_8 \;exp\;[+ i \; (k_2 - k_3 + k_4) \; y]\Bigr),
\end{eqnarray}  
where $k_2, k_3$ and $k_4$ are the momenta associated with the mode expansions of 
$C(y), \psi^\dagger (y)$ and $\psi(y)$, respectively. The 
operators $a_i$ $(i = 1, 2, 3,....8)$ are
\begin{eqnarray} 
&& a_1 = c(k_2) \; (b^\gamma)^\dagger (k_3) \; (u^\gamma)^\dagger (k_3) \; b^\sigma (k_4) \; 
u^\sigma (k_4),\nonumber\\ 
&& a_2 = c(k_2) \; (b^\gamma)^\dagger (k_3) \; (u^\gamma)^\dagger (k_3)\; 
(d^\sigma)^\dagger (k_4) \; v^\sigma (k_4),\nonumber\\
&& a_3 = c(k_2)\;  d^\gamma (k_3) \; (v^\gamma)^\dagger (k_3) 
\; b^\sigma (k_4) \; u^\sigma (k_4), \nonumber\\
&& a_4 = c(k_2) \; d^\gamma (k_3) \; (v^\gamma)^\dagger (k_3) \; (d^\sigma)^\dagger (k_4) \;
v^\sigma (k_4),\nonumber\\ 
&& a_5 = c^\dagger(k_2) \; (b^\gamma)^\dagger (k_3) \; (u^\gamma)^\dagger (k_3) \;
b^\sigma (k_4) \; u^\sigma (k_4),\nonumber\\ 
&& a_6 = c^\dagger(k_2) \; (b^\gamma)^\dagger (k_3) \; (u^\gamma)^\dagger (k_3) \;
(d^\sigma)^\dagger (k_4) \; v^\sigma (k_4),\nonumber\\ 
&& a_7 = c^\dagger(k_2) \; d^\gamma (k_3) \; (v^\gamma)^\dagger (k_3) \; b^\sigma (k_4) \;
u^\sigma (k_4),\nonumber\\ 
&& a_8 = c^\dagger(k_2)\;  d^\gamma (k_3) \; (v^\gamma)^\dagger (k_3) \; 
(d^\sigma)^\dagger (k_4) \; v^\sigma (k_4).       
\end{eqnarray}
It will be noted that we have {\it not} written operators $a_i$ in the normal 
ordered form and maintained the order as they have appeared (because the relevant part of 
$Q_b$ is still not in the quadratic form).  
Ultimately, we have to compute the anticommutator between the eight terms of the 
relevant part of $Q_b$ and the two terms of the mode expansion of $\psi (x, t)$ [cf. (15)]
that are written in terms of operators $b^\alpha (k)$ and $(d^\alpha)^\dagger (k).$

It is clear, from the expansion of $\psi (x, t)$ and the above eight terms, that there would, 
in totality, be sixteen anticommutators in the computation of $\{ \psi , Q_b \}$. At 
this stage, the fermionic (i.e. $\psi^2 (x, t) = 0, (\psi^\dagger)^2 (x, t) = 0 $) nature 
of the fields $\psi (x, t)$ and $\psi^\dagger (x, t)$ helps us immensely. For instance, it 
can be seen that the following relationships 
amongst the creation and annihilation operators 
ensue from the first condition $\psi^2 (x, t) = 0$,
namely;
\begin{eqnarray} 
&& \psi^2 (x, t) = \frac {1}{2} \; \{ \psi (x, t), \; \psi (x, t)\} = 0 \; \Longrightarrow  
\{ b^\alpha (k), \;  b^\beta (k^\prime )\} = 0, \nonumber\\
&& \{(d^\alpha)^\dagger (k), \; (d^\beta )^\dagger (k^\prime)\} = 0, \qquad 
\{ b^\alpha (k), \; (d^\beta )^\dagger (k^\prime ) \} = 0.  
\end{eqnarray} 
Similarly $(\psi^\dagger)^2 (x, t) = 0$ implies the following
\begin{eqnarray} 
\fl \quad \bigl \{(b^\alpha)^\dagger (k), \; (b^\beta )^\dagger (k^\prime ) \bigr \} = 0, \quad 
\bigl \{(b^\alpha)^\dagger (k), \; d^\beta (k^\prime) \bigr \} = 0, \quad 
\bigl \{ d^\alpha (k), \; d^\beta (k^\prime ) \bigr \} = 0.  
\end{eqnarray}
Note that the relations (24) and (25) are also valid for $k = k^\prime$ (because of 
the limiting case). The above canonical brackets help us in evaluating sixteen 
anticommutators in a simple manner because many of them vanish.

It is straightforward to check that the first term of the mode expansion of 
$\psi (x, t)$ [cf. (15)] and the first term of the expansion (22) lead to 
the following anticommutator (for $ t = 0 $), namely;
\begin{eqnarray}
\sum_\beta \sum_\gamma \sum_\sigma \int dy \frac {dk_1 \; dk_2 \; dk_3 \; dk_4}{(2\pi )^2 
\surd {(2k^0_1 \; 2k^0_2 \;  2k^0_3 \;  2k^0_4)}} \; \nonumber\\
\Bigl \{ b^\beta (k_1) \; u^\beta (k_1) \; e^{- i k_1 \; x}, \; 
a_1 \; e^{- i \; (k_2 - k_3 + k_4) \; y} \Bigr \}, 
\end{eqnarray}    
where $k_1, \; k_2, \; k_3, \; k_4$ are the momenta corresponding to the fields 
$\psi (x), \; C (y)$, $\psi^\dagger (y)$ and $\psi (y)$,
respectively, in the phase space. It is clear from the expression for $``a_1"$ 
[cf. (23)] and the relevant anticommutator of (24) that we have 
\begin{eqnarray}
\fl \qquad && - \sum_\beta \sum_\gamma \sum_\sigma \int dy \frac {dk_1 \; dk_2 \; dk_3 \; 
dk_4}{(2\pi )^2 \surd {(2k^0_1 \; 2k^0_2 \; 2k^0_3 \; 2k^0_4)}} \; c (k_2) \; \bigl \{b^\beta (k_1), 
(b^\gamma)^\dagger (k_3) \bigr \} \nonumber\\
\fl \qquad && u^\beta (k_1) \; (u^\gamma )^\dagger (k_3) \; b^\sigma (k_4) \; u^\sigma (k_4) \; 
e^{- i \; k_1 \;x} \; e^{ - i\; (k_2 - k_3 + k_4) \; y}. 
\end{eqnarray}
Here a couple of points are to be noted. First, since we are exploiting here the equal-time 
anticommutators, all the field expansions have been written for $t = 0$. Second, since the
(anti-)ghost fields are decoupled from the rest of the theory, the  
annihilation and creation operators $c (k_2)$ and $ c^\dagger (k_2)$ anticommute with the rest 
of the fermionic creation and annihilation operators. This can also be verified by the fact that,
under the ghost transformations, the Dirac field $\psi$ does {\it not} transform
($ s_g \psi = i \; [ \psi, Q_g ] = 0$). Taking the help of $Q_g$ from (10) and $\psi (x, t)$ 
from (15), it is clear that $ \{ b^\beta (k_1), \; c(k_2) \} = 0, 
\{ b^\beta (k_1), \; c^\dagger (k_2) \} = 0, \{ (d^\beta)^\dagger (k_1), \; c(k_2) \} = 0, 
\{ (d^\beta)^\dagger (k_1), \; c^\dagger (k_2) \} = 0$, etc., where we have used the expression for $Q_g$
when it is expressed in terms of the creation and annihilation operators as:
$Q_g = - \int dk \big[\bar c^\dagger(k)\, c(k) + c^\dagger(k)\, \bar c(k) \big]$ (see, e.g. [3]).

Comparing the above exponential with the exponential of the first term [cf. (19)] of the l.h.s. 
(i.e. $C (x) \psi (x)$), it is straightforward that if we choose
\begin{eqnarray}
\{ b^\beta (k_1), \;(b^\gamma )^\dagger (k_3)\} = - \;\delta^{\beta \gamma} 
\; \delta (k_1 - k_3),
\end{eqnarray}
we can match the exponential of the first term of the l.h.s. [cf. (19)] if 
$k_2 \rightarrow k$ and $k_4 \rightarrow k^\prime$ because we have the following 
definition of the Dirac $\delta$-function in the space part of the 2D spacetime manifold, namely;
\begin{eqnarray}
\int \frac {dk_1}{(2\pi)} \; e^{- i \;(k_1 \; x \; - \; k_3 \; y)}\Big|_{k_1 = k_3} 
= \; \delta (x -y).
\end{eqnarray}
With input from (28), we obtain explicitly the following expression 
\begin{eqnarray}
\fl \qquad && \sum_\beta \sum_\gamma \sum_\sigma \int dy \frac {dk_1 \; dk_2 \; dk_3 \; 
dk_4}{(2\pi )^2 \surd {(2k^0_1 \; 2k^0_2 \; 2k^0_3 \; 2k^0_4)}} \; c (k_2) \; 
\delta^{\beta \gamma} \; \delta (k_1 - k_3) \; u^\beta (k_1) \nonumber\\
\fl \qquad && (u^\gamma )^\dagger (k_3) \;
b^\sigma (k_4) \; u^\sigma (k_4) \; e^{-i\; k_1\; x} \;e^{- \; i \;(k_2 - k_3 + k_4) \;y}\nonumber\\
\fl \qquad && = \sum_\sigma \int dy \frac {dk_1 \; dk_2 \; dk_4}{(2\pi )^2 \; 2k^0_1 
\surd {(2k^0_2 \; 2k^0_4)}} \; 
e^{- i \;k_1 \; (x - y)} \; e^{- i \;(k_2 + k_4) \; y} \; c(k_2) \nonumber\\
\fl \qquad && \Bigl ( \sum_\beta u^\beta (k_1) \; {\bar u}^\beta (k_1) \Bigr )
\; \gamma^0 \; b^\sigma (k_4) \; u^\sigma (k_4),
\end{eqnarray}
where we have inserted $(\gamma^0)^2 = I$ appropriately at a suitable place.

The above type of expression also emerges from the anticommutator of the second term of 
$\psi (x, t)$ [cf. (15)] with the third term of (22), namely;  
\begin{eqnarray}
\fl \qquad && \sum_\beta \sum_\gamma \sum_\sigma \int dy \frac {dk_1 \; dk_2 \; dk_3 \; dk_4}{(2\pi )^2 
\surd {(2k^0_1 \; 2k^0_2 \; 2k^0_3 \; 2k^0_4)}} \; \nonumber\\
\fl \qquad &&\{(d^\beta)^\dagger (k_1) \; v^\beta (k_1) \; e^{+ i k_1 \; x}, \;a_3 \; 
e^{- i (k_2 + k_3 + k_4) \;y)} \} \nonumber\\ 
\fl \qquad && = -  \sum_\beta \sum_\gamma \sum_\sigma 
\int dy \frac {dk_1 \; dk_2 \; dk_3 \; dk_4}{(2\pi )^2 
\surd {(2k^0_1 \; 2k^0_2 \;  2k^0_3 \; 2k^0_4)}} \;
c (k_2)\; \{(d^\beta)^\dagger (k_1), \; d^\gamma (k_3)\} \nonumber\\
\fl \qquad && v^\beta (k_1) \; (v^\gamma )^\dagger (k_3) \; b^\sigma (k_4) \; u^\sigma (k_4) \; 
e^{i \; k_1 \; x} \; e^{- \; i \;(k_2 + k_3 + k_4) \;  y},    
\end{eqnarray} 
due to the fact that $\{(d^\beta)^\dagger (k_1), \; b^\sigma (k_4) \} \; = \; 0$ 
[cf. (24)] and $c(k_2)$ does anticommute with the operator $(d^\beta)^\dagger (k_1)$.

Finally, we obtain the analogue of equation (30) as 
\begin{eqnarray}  
\sum_\sigma \int dy \frac {dk_1 \; dk_2 \; dk_4}{(2\pi )^2 \; 2k^0_1 \surd {(2k^0_2 \; 2k^0_4)}} 
\; e^{+ i \; k_1 \; (x - y)} e^{- i \; (k_2 + k_4) \; y} \; c(k_2) \nonumber\\
\Bigl ( \sum_\beta v^\beta (k_1)  \;  {\bar v}^\beta (k_1) \Bigr )
\; \gamma^0 \; b^\sigma (k_4) \; u^\sigma (k_4),
\end{eqnarray}
if we exploit the anticommutator 
$\{(d^\beta)^\dagger (k_1), \; d^\gamma (k_3) \} \; 
= \;- \; \delta^{\beta \gamma} \; \delta (k_1 - k_3)$.
Exploiting the trick of our Appendix A and using 
equation (11), it can be checked that the sum of (30) and (32) leads to
\begin{eqnarray}
&& \sum_\sigma \int \frac{dk_2 \; dk_4}{(2\pi) \surd{(2k^0_2 \; 2k^0_4)}}\; 
e^{- i \; (k_2 + k_4) \; x} \; c(k_2) \; b^\sigma (k_4) \; u^\sigma(k_4),
\end{eqnarray} 
which is the first term (modulo  ``e") of the l.h.s. of (19) in the limit
$k_2 \to k, \; k_4 \to k^\prime$ and  $\sigma \to \alpha$. In exactly similar fashion,
if we use (62) (see below), it can be checked that the {\it sum} of the anticommutators between

(i) the {\it first} term of $\psi (x, t)$ and the {\it second} term of (22) and that of the 
{\it second} term of $\psi (x, t)$ with the {\it fourth} term of (22) yields the second term of 
the l.h.s. [cf. equation (19)],

(ii) the {\it first} term of $\psi (x, t)$ and the {\it fifth} term of (22) and that of the 
{\it second} term of $\psi (x, t)$ with the  {\it seventh} term of (22) produces the third term of 
the l.h.s. [cf. equation (19)], and

(iii) the {\it first} term of $\psi (x, t)$ and the {\it sixth} term of (22) and that of the 
{\it second} term of $\psi (x, t)$ with the {\it eighth} term of (22) leads to the fourth term of 
the l.h.s. [cf. equation (19)].

We lay emphasis on the fact that it is the comparison of the exponentials from the l.h.s. 
and r.h.s. of (18) that dictates the non-vanishing brackets to be (62) (see, Sec. 6 below). 
Rest of the anticommutators amongst the fermionic creation and annihilation operators 
turn out to be zero.  Exactly similar kind of computations, with the anti-BRST charge
$Q_{ab}$, produces the basic canonical brackets to be (62) (see below).
We conclude that symmetry transformations (and their generators) of a Hodge theory 
can {\it replace} the mathematical definition of the canonical conjugate momenta.

\section{(Anti-)co-BRST symmetries: basic brackets}

The Lagrangian density (2) also respects the nilpotent (anti-)co-BRST symmetry 
transformations $s_{(a)d}$ in the massless ($m = 0$) limit of the Dirac fields. 
Form the (anti-)co-BRST symmetry transformations (5) and the corresponding charges (6), 
the canonical brackets amongst the creation and annihilation operators
can be computed.

The non--vanishing brackets for the free theory (without matter field) has 
already been derived in our previous work [3]. We shall focus, therefore, only 
on the derivation of the canonical brackets for the matter fields of 2D QED.
Once again, as has 
been done with (anti-)BRST charges, the decisive role is played by the definition
of the generator ($Q_{(a)d}$) of the continuous symmetries ($s_{(a)d}$) 
(i.e. $s_{(a)d} \psi  = - i \{\psi , Q_{(a)d} \}$). Let us take an example for the 
sake of clarification, namely; 
\begin{eqnarray}
s_d \; \psi = - \; i \; \{\psi ,\;  Q_d \} = \; 
- \; i \; e \; \bar C \; \gamma_5 \; \psi \Longrightarrow 
\{\psi ,\;  Q_d \} = e \; \bar C \; \gamma_5 \; \psi.
\end{eqnarray}
The term that would contribute from $Q_d$ [cf. (6)] is
\begin{eqnarray}
Q^{(R)}_d \; = \; e \int dx \; \psi^\dagger \; \gamma_0 \gamma_1 \; \bar C \; \psi \; = \;  
- \; e \int dx \; \psi^\dagger \; \gamma_5 \; \bar C \; \psi,
\end{eqnarray}
for the determination of the continuous symmetry transformation connected with $s_d$. 
Here the superscript $(R)$ on the conserved charge $Q_d$ (i.e. $Q^{(R)}_d$) denotes the relevant part of the 
exact expression for $Q_d$ which contributes in our computation.
Similarly, one can write all the rest of the fermionic transformations $s_{(a)d}$ 
in terms of the corresponding generators. Following the tricks developed in the 
case of the nilpotent (anti-)BRST symmetry transformations, it can be checked that exactly 
the same non--vanishing [cf. (62)] anticommutators (amongst the creation and annihilation 
operators) emerge from this exercise, too.

To corroborate the above statement,  we provide some of the key steps that are needed in the determination of the
anticommutators (62) from the continuous symmetry transformations $s_{(a)d}$ generated by the
(anti-)co-BRST charges $Q_{(a)d}$. It is evident from equations (34) and (35) that
we have, for the specific case of co-BRST transformation ($s_d \psi = - i e \gamma_5 \bar C 
\psi = - i \{\psi, Q_d \}$), the following expression:
\begin{eqnarray}
e \;\bar C\; \gamma_5 \psi = + e \;\gamma_5 \;{\displaystyle \int} \; dy\;
\bigl \{ \psi (x), \;\bar C (y)\, \psi^\dagger (y)\, \psi (y) \bigr \},
\end{eqnarray} 
where we have used $\bar C \psi^\dagger + \psi^\dagger \bar C = 0$. The above expression,
finally, reduces to the following straightforward equality:
\begin{eqnarray}
\bar C (x)\; \psi (x) =  {\displaystyle \int} \; dy\;
\bigl \{ \psi (x), \;\bar C (y)\, \psi^\dagger (y)\, \psi (y) \bigr \},
\end{eqnarray}
which is similar in appearance as the corresponding 
nilpotent BRST transformations [cf. (18), (21)] with the 
replacement $C \to \bar C$. In fact, the expression in (37) is exactly same as 
the one connected with the anti-BRST symmetry transformations and 
their corresponding generator (i.e. anti-BRST charge).

The l.h.s. of (37) (for $ t = 0$) can be written, analogous to (19), as
\begin{eqnarray}
\fl \qquad && \int \frac { d k \; d k^\prime}{ (2 \pi ) \surd (2 k_0 \; 2 k^\prime_0 )}
\sum_\alpha \Bigl( \bar c(k) \; b^\alpha (k^\prime) \; u^\alpha (k^\prime)\; 
exp \;[- i \; (k + k^\prime ) \; x]\nonumber\\ 
\fl \qquad && + \bar c (k) \; (d^\alpha)^\dagger (k^\prime) \; v^\alpha (k^\prime) \; 
exp \;[- i \; (k - k^\prime ) \; 
 x] + \bar c^\dagger (k) \; b^\alpha (k^\prime)\;  u^\alpha (k^\prime) \nonumber\\
\fl \qquad && exp \;[ + i \; (k - k^\prime ) \; x] + \bar c^\dagger (k) \; (d^\alpha)^\dagger (k^\prime) 
\; v^\alpha (k^\prime) \; exp \;[ + i \; (k + k^\prime ) \; x] \Bigr).
\end{eqnarray} 
Similarly, the r.h.s. of (37) can be written, analogous to (22), as
\begin{eqnarray}
\fl \qquad && \sum_\gamma \sum_\sigma \int dy \int \frac{dk_2 \; dk_3 \; dk_4}{(2\pi )^{3/2} 
\surd {(2k^0_2 \; 2k^0_3 \; 2k^0_4)}} \Bigl( b_1 \;
exp \;[- i \; (k_2 - k_3 + k_4) \; y] \nonumber\\
\fl \qquad && + \; b_2 \;exp\;[- i \; (k_2 - k_3 - k_4) \; y] + \; b_3 \;
exp \;[- i \; (k_2 + k_3 + k_4) \; y] \nonumber\\
\fl \qquad && + \; b_4 \;exp\;[- i \; (k_2 + k_3 - k_4) \; y] + \; b_5 \;
exp \;[+ i \; (k_2 + k_3 - k_4) \; y] \nonumber\\
\fl \qquad && + \; b_6 \;exp\;[+ i \; (k_2 + k_3 + k_4) \; y] + \; b_7 \;
exp\;[+ i \; (k_2 - k_3 - k_4) \; y] \nonumber\\
\fl \qquad && + \; b_8 \;exp\;[+ i \; (k_2 - k_3 + k_4) \; y]\Bigr),
\end{eqnarray}  
where we have taken $t = 0$ for the relevant portion of the $Q_d$ because it is 
a conserved quantity. Furthermore, we are using here the equal-time anticommutators
which enforce us to take the mode expansion of $\psi (x, t)$ at $t = 0$, too. The momenta $k_2, k_3$ 
and $k_4$, in the above, are associated with the mode expansions of 
$\bar C(y), \psi^\dagger (y)$ and $\psi(y)$, respectively. The 
explicit form of the operators $b_i\; (i = 1, 2, 3,....8)$, in the above,  are
\begin{eqnarray} 
&& b_1 = \bar c(k_2) \; (b^\gamma)^\dagger (k_3) \; 
(u^\gamma)^\dagger (k_3) \; b^\sigma (k_4) \; 
u^\sigma (k_4),\nonumber\\ 
&& b_2 = \bar c(k_2) \; (b^\gamma)^\dagger (k_3) \; (u^\gamma)^\dagger (k_3)\; 
(d^\sigma)^\dagger (k_4) \; v^\sigma (k_4),\nonumber\\
&& b_3 = \bar c(k_2)\;  d^\gamma (k_3) \; (v^\gamma)^\dagger (k_3) 
\; b^\sigma (k_4) \; u^\sigma (k_4), \nonumber\\
&& b_4 = \bar c(k_2) \; d^\gamma (k_3) \; (v^\gamma)^\dagger (k_3) \; (d^\sigma)^\dagger (k_4) \;
v^\sigma (k_4),\nonumber\\ 
&& b_5 = \bar c^\dagger(k_2) \; (b^\gamma)^\dagger (k_3) \; (u^\gamma)^\dagger (k_3) \;
b^\sigma (k_4) \; u^\sigma (k_4),\nonumber\\ 
&& b_6 = \bar c^\dagger(k_2) \; (b^\gamma)^\dagger (k_3) \; (u^\gamma)^\dagger (k_3) \;
(d^\sigma)^\dagger (k_4) \; v^\sigma (k_4),\nonumber\\ 
&& b_7 = \bar c^\dagger(k_2) \; d^\gamma (k_3) \; (v^\gamma)^\dagger (k_3) \; b^\sigma (k_4) \;
u^\sigma (k_4),\nonumber\\ 
&& b_8 = \bar c^\dagger(k_2)\;  d^\gamma (k_3) \; (v^\gamma)^\dagger (k_3) \; 
(d^\sigma)^\dagger (k_4) \; v^\sigma (k_4).       
\end{eqnarray} 
It should be noted that we have not yet written the operators $b_i$ in the normal ordered form and we have
maintained the order as they appear in the expression for a portion of $Q_d$
(where the local fields $ \bar C (y), \psi^\dagger (y) $ and $\psi (y)$ are present).

Exploiting the inputs from (24) and the fact that the ghost transformation for
the field $\psi (x)$ is zero (i.e. $s_g \psi = + i [\psi, Q_g] = 0$), it is clear,
from the expression $Q_g = i\,\int dx\; [C\, \dot {\bar C} +  \bar C\,\dot C] \equiv 
 - \int dk \big[\bar c^\dagger(k)\, c(k) + c^\dagger(k)\, \bar c(k) \big]$ [cf. (10)]
and expansion in (15) (for the field $\psi (x)$),
that the following anticommutators are true, namely;
\begin{eqnarray}
\fl \qquad &&\{b^\beta (k_1),\, c(k_2) \} = 0, \quad \; \{b^\beta (k_1),\, c^\dagger (k_2) \} = 0, \quad \quad
\{b^\beta (k_1), \bar c(k_2) \} = 0, \nonumber\\       
\fl \qquad && \{b^\beta (k_1),\, \bar c^\dagger (k_2) \} = 0, \quad \{(d^\beta)^\dagger (k_1),\, c(k_2) \}
= 0, \quad   \{(d^\beta)^\dagger (k_1),\, c^\dagger (k_2) \} = 0, \nonumber\\
\fl \qquad && \{(d^\beta)^\dagger (k_1), \,\bar c(k_2) \} = 0, \qquad
\{(d^\beta)^\dagger (k_1),\, \bar c^\dagger (k_2) \} = 0.
\end{eqnarray}
We note that all the arguments of the BRST transformations
(i.e. $s_b \psi = - i \{\psi,\, Q_b \} = - \,i\, e\, C\, \psi$), discussed 
in the main body of our previous section, would be applicable in our present discussion (connected
with the continuous symmetry transformation $s_d \psi = - i \{ \psi,\, Q_d \}
= - i\, e\, \gamma_5 \,\bar C\, \psi$) as well. It is obvious now that we shall obtain, ultimately,
the non--vanishing brackets as (62) and the rest of the brackets would turn out to be zero.
Similar kind of computations can be performed with $Q_{ad}$ which will, once again, 
lead to the derivation of the non-vanishing canonical brackets (62).
We would like to lay emphasis on the fact that the normal ordering in (40)
 would not affect the main results of our analysis. In other words, the 
non--vanishing brackets (62) would remain unaffected by normal ordering in equation (40).

\section{Unique bosonic symmetry: basic brackets}

We have a unique bosonic symmetry ($s_\omega$) in our theory. The generator ($Q_\omega$)
of the transformations has been quoted in Sec. 2 [cf. (8)].
This bosonic charge $Q_\omega$
generates the bosonic transformations $s_\omega$ on the fermionic fields as [8,9]
\begin{eqnarray}
\fl \qquad s_\omega \psi = - \; i \; e \big [\gamma_5 (\partial \cdot A) + E \big ]  \psi, \qquad 
s_\omega \bar \psi = +\, i \; e \, \bar \psi \,\big [E - \gamma_5 (\partial \cdot A) \big ],
\end{eqnarray}
which constitute the symmetry 
invariance of the Lagrangian density (2). The key equation [consistent with (13)]
that leads 
to the determination of the canonical brackets amongst the creation and 
annihilation operators is the following relationship:
\begin{eqnarray}
s_\omega \psi \; = \;   + \; i \; \big [ \psi, Q_\omega \big ] \; = \;  
- \; i \; e \; \big [\gamma_5 (\partial \cdot A) + E \big ] \psi, \nonumber\\ 
s_\omega \bar \psi \; = \;  + \; i \; \big [ \bar \psi, Q_\omega \big ] \; = \;  
+ \, i \, e \, \bar \psi\, \big [E - \gamma_5 (\partial \cdot A)\big ].  
\end{eqnarray}
It is evident that, finally, we have to compute $  [\psi, Q_\omega ]  
=  - e \; \big [\gamma_5 (\partial \cdot A) + E \big ] \psi$ and 
$  [ \bar \psi, Q_\omega ]  =  + \; e \; \bar \psi \,\big [E - \gamma_5 
(\partial \cdot A)\big ]$ for the determination of the 
canonical anticommutators amongst the creation and annihilation operators.
For this purpose, the portion of the bosonic charge ($Q_\omega$) that would 
contribute is: $Q^{(R)}_\omega = e \int dx \;  [ (\partial \cdot A)
\psi^\dagger \; \gamma_5 \; \psi + E \; \psi^\dagger \; \psi ]$ (where 
$(\gamma_0)^2 = I, \gamma_5 = \gamma^0 \gamma^1$ are used).
Here the superscript $(R)$ denotes the relevant part of $Q_\omega$ that contribute 
in the computation of $s_\omega \, \psi$ and $s_\omega \, \bar \psi$.
It is to be emphasized that, for the free theory (without matter fields), the basic 
canonical brackets have been determined in our previous work [3]. This is why, 
we have concentrated only on the determination of brackets for the matter fields 
from their symmetry transformations.

Once again, applying  the rules of the commutators and anticommutators, we end up 
with the non-vanishing anticommutators as (62) (see below) that have been 
derived by canonical method. Some of the key steps of our present exercise 
are illustrated here for the readers' convenience. It is evident from (43) that
\begin{eqnarray}
- \; e \; [ \gamma_5 (\partial \cdot A) + E ] \; \psi \; = \; [ \psi, \; Q_\omega ],
\end{eqnarray} 
where the term that contributes from $Q_\omega$ is: $ e \int dy\; [(\partial \cdot A)\,
\gamma_5\, \psi^\dagger \,\psi + E \,\psi^\dagger\, \psi ]$. Thus, the factor $``e"$ cancels
from l.h.s. and r.h.s. of equation (44). Plugging in the expansions from (12) (with $\partial \cdot A
= \partial_\mu A^\mu $, $ E = - \varepsilon ^{\mu \nu} \partial _\mu A_\nu $) and 
(15), we obtain the l.h.s. of (44) (for $t = 0$) as follows: 
\begin{eqnarray}
\fl \qquad && i \int \; \frac{d k \; d k^\prime}{(2 \pi) \surd {(2k_0 \; 2k^\prime_0)}}
\; (\varepsilon^{\mu \nu} \; k_\nu + \gamma_5 \; k^\mu)\nonumber\\
\fl \qquad && \sum_\alpha \Big[ a^\dagger_\mu (k) \; u^\alpha (k^\prime) \; b^\alpha (k^\prime)
e^{ + \; i (k - k^\prime) \; x } + a^\dagger_\mu (k) \; (d^\alpha)^\dagger (k^\prime) 
\; v^\alpha (k^\prime)  \; e^{ + \; i (k + k^\prime) \; x } \nonumber\\
\fl \qquad && - a_\mu (k) \; u^\alpha (k^\prime) \; b^\alpha (k^\prime)
e^{ - \; i (k + k^\prime) \; x } - a_\mu (k) \; (d^\alpha)^\dagger (k^\prime) 
\; v^\alpha (k^\prime)  \; e^{ -  i (k - k^\prime)  x } \Big ],
\end{eqnarray}
where $k$ and $k^\prime$ are the momenta in the phase space corresponding 
to the field expansions in (12) (for $A_\mu (x)$ field) and (15) (for $\psi(x)$ field).

The relevant expression for $Q_\omega$ on the r.h.s. is: 
$\int dy\; \bigl ( [(\partial \cdot A) (y) \gamma_5 + E (y)] \; \psi^\dagger (y) \psi (y) \bigr )$.
Exploiting the expansions of (12), (15) and (16) at $ t = 0 $, we obtain the 
following expression for the relevant $Q_\omega$ operator, namely;
\begin{eqnarray}
\fl \qquad && - i \; \sum_\gamma \sum_\sigma \int \; d y \frac{d k_2 \; d k_3 \; d k _4}
{(2 \pi)^{3/2} \surd {( 2k^0_2 \; 2k^0_3 \; 2k^0_4)}} \; (\varepsilon^{\mu \nu} 
\; k_{2 \nu} + \gamma_5 \; k^\mu_2) \nonumber\\
\fl \qquad && \Big [ c^{(1)}_\mu \; exp \; ( + \;  i \; (k_2 + k_3 - k_4) \; y ) + \; c^{(2)}_\mu \; 
exp \;( + \;  i \; (k_2 + k_3 + k_4) \; y ) \nonumber\\
\fl \qquad && + \; c^{(3)}_\mu \; exp \; ( + \;  i \; (k_2 - k_3 - k_4) \; y ) + \; c^{(4)}_\mu \;
 exp \; ( + \;  i \; (k_2 - k_3 + k_4) \; y ) \nonumber\\ 
\fl \qquad && + \; c^{(5)}_\mu \; exp \; ( - \;  i \; (k_2 - k_3 + k_4) \; y ) + \; c^{(6)}_\mu \; 
exp \;( - \;  i \; (k_2 - k_3 - k_4) \; y )  \nonumber\\
\fl \qquad && +  c^{(7)}_\mu \; exp \; ( -   i \; (k_2 + k_3 + k_4) \; y ) +  c^{(8)}_\mu \; 
exp \; ( -   i \; (k_2 + k_3 - k_4) \; y ) \Big ],
\end{eqnarray}
where $k_2, k_3$ and $k_4$ are the momenta corresponding to the fields $A_\mu (y)$, 
$\psi^\dagger (y)$ and $\psi (y)$, respectively. The operators
$c^{(i)}_\mu \; ( i = 1, 2, 3, .... 8)$ are 
\begin{eqnarray}
&& c^{(1)}_\mu = a^\dagger_\mu (k_2) \; (b^\gamma)^\dagger (k_3) \; (u^\gamma)^\dagger (k_3) \; b^\sigma (k_4) \; 
u^\sigma (k_4)\nonumber,\\ 
&& c^{(2)}_\mu = a^\dagger_\mu (k_2) \; (b^\gamma)^\dagger (k_3) \; (u^\gamma)^\dagger (k_3)\; 
(d^\sigma)^\dagger (k_4) \; v^\sigma (k_4),\nonumber\\
&& c^{(3)}_\mu = a^\dagger_\mu (k_2)\;  d^\gamma (k_3) \; (v^\gamma)^\dagger (k_3) 
\; b^\sigma (k_4) \; u^\sigma (k_4), \nonumber\\
&& c^{(4)}_\mu = a^\dagger_\mu (k_2) \; d^\gamma (k_3) \; (v^\gamma)^\dagger (k_3) \; (d^\sigma)^\dagger (k_4) \;
v^\sigma (k_4),\nonumber\\ 
&& c^{(5)}_\mu = - \; a_\mu (k_2) \; (b^\gamma)^\dagger (k_3) \; (u^\gamma)^\dagger (k_3) \;
b^\sigma (k_4) \; u^\sigma (k_4),\nonumber\\ 
&& c^{(6)}_\mu = - \; a_\mu(k_2) \; (b^\gamma)^\dagger (k_3) \; (u^\gamma)^\dagger (k_3) \;
(d^\sigma)^\dagger (k_4) \; v^\sigma (k_4),\nonumber\\ 
&& c^{(7)}_\mu = - \; a_\mu (k_2) \; d^\gamma (k_3) \; (v^\gamma)^\dagger (k_3) \; b^\sigma (k_4) \;
u^\sigma (k_4),\nonumber\\ 
&& c^{(8)}_\mu = - \; a_\mu (k_2)\;  d^\gamma (k_3) \; (v^\gamma)^\dagger (k_3) \; 
(d^\sigma)^\dagger (k_4) \; v^\sigma (k_4). 
\end{eqnarray}
where the order of the creation and annihilation operators has been maintained (without
any kind of implementation of the normal ordering).

Now the stage is set for the explicit computation of the bracket 
$ [ \psi, Q_\omega ] $. From expansion of $\psi (x)$ in (15), it is evident
that there would be sixteen commutators but many of them would vanish due 
to our earlier arguments. It can be checked that commutator of the 
first term of expansion of $\psi (x)$ [in (15)] and the first term of the 
relevant part of $Q_\omega$ in (46) (for $t = 0$) yields:  
\begin{eqnarray}
\fl \qquad && - i \; \sum_\beta \sum_\gamma \sum_\sigma \int \; d y \frac{d k_1 \; d k_2 \; d k_3 \; d k _4}
{(2 \pi)^2 \surd {( 2k^0_1 \; 2k^0_2 \; 2k^0_3 \; 2k^0_4)}} \; (\varepsilon^{\mu \nu} 
\; k_{2 \nu} + \gamma_5 \; k^\mu_2) \nonumber\\
\fl \qquad && \Big [ b^\beta (k_1), \; a^\dagger_\mu (k_2) \; (b^\gamma)^\dagger (k_3)  
\; b^\sigma (k_4) \Big ] \; u^\beta (k_1) \; (u^\gamma)^\dagger (k_3) \; u^\sigma (k_4)\nonumber\\
\fl \qquad && exp \; ({- i \; k_1 \; x}) \; exp \; ({ + i \; (k_2 + k_3 - k_4) \; y }).
\end{eqnarray}
Exploiting the appropriate rules of the commutators we obtain, ultimately, the 
following existing bracket, namely;
\begin{eqnarray}
\fl \qquad && - i \; \sum_\beta \sum_\gamma \sum_\sigma \int \; d y \frac{d k_1 \; d k_2 \; d k_3 \; d k _4}
{(2 \pi)^2 \surd {( 2k^0_1 \; 2k^0_2 \; 2k^0_3 \; 2k^0_4)}} \; (\varepsilon^{\mu \nu} 
\; k_{2 \nu} + \gamma_5 \; k^\mu_2) \nonumber\\
\fl \qquad && \; a^\dagger_\mu (k_2) \; \Big \{ b^\beta (k_1),  \; (b^\gamma)^\dagger (k_3)  \Big \}
 \; b^\sigma (k_4) \; u^\beta (k_1) \; (u^\gamma)^\dagger (k_3) \; u^\sigma (k_4)\nonumber\\
\fl \qquad && exp \; \bigl [- i \; k_1 \; x +  i \; (k_2 + k_3 - k_4) \; y  \bigr ],
\end{eqnarray}
where we have already exploited (24) and the commutator $ [ b^\beta (k_1), \; 
a^\dagger_\mu (k_2) ] = 0 $ due to the fact that there is no explicit mixing between the fields
$A_\mu (x)$ and $\psi (x)$ as far as our basic continuous symmetry transformations
$s_{(a)b}$ and $s_{(a)d}$ are concerned. Furthermore, under the ghost
transformations, fermionic fields $\psi$ and $\psi^\dagger$ do not
transform at all. Hence, there is no mixing here as well.

The exponentials of the first term of the l.h.s. of (44)
[which is explicitly expressed as (45)] and the above commutator would match if we 
exploit the relevant canonical bracket of (62) 
[i.e. $\{ b^\beta (k_1), (b^\gamma)^\dagger (k_3) \} = - \delta^{\beta\gamma} \delta (k_1 - k_3)$].
As a consequence, we obtain the following explicit expression:
\begin{eqnarray}
\fl \qquad i \sum_\sigma \; \int\; dy \frac{d k_1 \; d k_2 \; d k_3 \; d k _4}
{(2 \pi)^2 \surd {( 2k^0_1 \; 2k^0_2 \; 2k^0_3 \; 2k^0_4)}} \; (\varepsilon^{\mu \nu} 
\; k_{2 \nu} + \gamma_5 \; k^\mu_2) \; \delta (k_1 - k_3) \nonumber\\
\fl \qquad a^\dagger_\mu (k_2) \; \sum_\beta \Bigl( u^\beta (k_1) \; \bar u^\beta (k_1) \Bigr) \;
\gamma^0 \; b^\sigma (k_4) \; u^\sigma (k_4) \; e^{- i \; k_1 \; (x - y)} \; e^ { + i \; (k_2 - k_4) \; y }.
\end{eqnarray}
Similar kind of terms would be generated from the commutator of the {\it second}
term of $\psi$ [cf. (15)] and the {\it third} term of the relevant portion of
$Q_\omega$ [cf. (46)]. This can be expressed as
\begin{eqnarray}
\fl \qquad i \sum_\sigma \; \int\; dy \frac{d k_1 \; d k_2 \; d k_3 \; d k _4}
{(2 \pi)^2 \surd {( 2k^0_1 \; 2k^0_2 \; 2k^0_3 \; 2k^0_4)}} \; (\varepsilon^{\mu \nu} 
\; k_{2 \nu} + \gamma_5 \; k^\mu_2) \; \delta (k_1 - k_3) \nonumber\\
\fl \qquad a^\dagger_\mu (k_2) \; \sum_\beta \Bigl( v^\beta (k_1) \; \bar v^\beta (k_1) \Bigr) \;
\gamma^0 \; b^\sigma (k_4) \; u^\sigma (k_4) \; e^{+ i \; k_1 \; (x - y)} \; 
e^ { + i \; (k_2 - k_4) \; y }.
\end{eqnarray}
The sum of (50) and (51) (with the help of (17), definition and properties of the 
Dirac $\delta$-function) produces the first term of the l.h.s. [cf. (45)] where
one has to make the replacements: $k_2 \to k, k_4 \to k^\prime$ and $\sigma \to \alpha$.

It is now straightforward to check that the rest of the terms of the l.h.s.
[cf. (45)] can also be produced with various combinations of commutators
from the first and second terms of $\psi$ [in (15)] and some appropriate terms of (46). 
It is crucial to note that, in all these
computations, the canonical brackets (62) play very decisive roles as they
are the root cause behind the emergence of the correct powers of the exponentials
on the r.h.s. In other words, we conclude that an
accurate comparison of the exponentials from the l.h.s. and r.h.s. [of (44)] leads
to the derivation of the canonical brackets amongst the creation and annihilations
operators of (62). We would like to lay emphasis on the fact that the form of (62) 
would remain unaffected even if we perform the normal ordering in (47).

We also very briefly comment, in this section,  on the (anti)commutators generated
by the continuous ghost symmetry transformations. 
We note that, under the ghost continuous symmetry transformations
[cf. (9)], all the physical fields ($A_\mu, \psi, \bar \psi$) remain unchanged
(i.e. $s_g A_\mu = s_g \psi = s_g \bar \psi = 0$). 
Using $Q_g = - \int dk \big[\bar c^\dagger(k)\, c(k) + c^\dagger(k)\, \bar c(k) \big]$ and (13), it is
elementary to check that the (anti-)ghost creation and annihilation operators
($c(k), c^\dagger (k), \bar c (k), \bar c^\dagger (k)$)
anticommute with such kind of fermionic operators that appear in the mode
expansions of $\psi$ and $\psi^\dagger$ (i.e. 
$b^\alpha (k), (b^\alpha)^\dagger (k), d^\alpha (k), (d^\alpha)^\dagger (k)$) 
and commute with the bosonic
operators (i.e. $a_\mu (k), a_\mu^\dagger (k)$) 
that appear in the normal mode expansion of the field $A_\mu$.
As evident from our earlier discussions, we have taken into account the anticommutativity property 
of $c(k)$, $c^\dagger(k)$, $\bar c(k)$, $\bar c^\dagger (k)$ with the creation and annihilation operators of 
$\psi$ and $\psi^\dagger$ and commutativity property with the operators  $a_\mu(k)$ and $a_\mu^\dagger (k)$ of the
bosonic gauge field $A_\mu$. The above properties are consistent with our argument.

\section{Canonical quantization scheme: Lagrangian formalism}
For the sake of completeness of our present work, we derive here the canonical brackets for {\it all}
the creation and annihilation operators of the interacting 2D model of Hodge theory (i.e 2D QED).    
It is evident that the canonical conjugate momenta from the Lagrangian density (2),
for the basic fields of the free (i.e. $\psi = \bar \psi = 0$) theory, are (see, e.g. [3] for details)
\begin{eqnarray}
\Pi^\mu = \frac {\partial {\cal L}_{(b)}}{\partial (\partial_0 A_\mu )} = - F^{0 \mu}
- \eta^{0 \mu} (\partial \cdot A),\nonumber\\
\Pi_C = \frac {\partial {\cal L}_{(b)}}{\partial (\partial_0 C)} = + \,i \,\dot {\bar C}, \;
\quad \Pi_{\bar C} = \frac {\partial {\cal L}_{(b)}}{\partial (\partial_0 \bar C)} 
= - \,i\, \dot C.
\end{eqnarray}
As a consequence, we have the following canonical 
commutator and anti-commutators for the theory in 2D, namely; 
\begin{eqnarray}
\fl \qquad && [A_\mu (x, t), \Pi_\nu (y, t)] = i \eta_{\mu\nu} \delta (x - y), \nonumber\\
\fl \qquad && \{\bar C(x, t), \Pi_{\bar C} (y, t) \} = i \delta (x - y) \Rightarrow 
\{\bar C(x, t), \dot{C} (y, t) \} = - \delta (x - y), \nonumber\\
\fl \qquad && \{C(x, t), \Pi_C (y, t) \} = i \delta (x - y) \Rightarrow 
\{ C(x, t), \dot{\bar C} (y, t) \} =  \delta (x - y). 
\end{eqnarray} 
All the rest of the brackets are zero. It is clear that here {\it two} of the main ingredients  
of the canonical quantization scheme have been exploited. These are the elevation of the 
(graded) Poisson brackets to the canonical (anti)commutators and the spin-statistics 
theorem. The top entry, in the above, implies the following commutators in terms of the 
components of the 2D gauge field $A_\mu$ and the corresponding conjugate momenta:
\begin{eqnarray}
&& [A_0 (x, t), \; (\partial \cdot A) (y, t) ] \; = \; - \; i \; \delta (x - y), \nonumber\\
&& [A_i (x, t), \; E_j (y, t) ] \; = \;  i \; \delta_{ij} \; \delta (x - y).
\end{eqnarray}
The above form of the commutators would turn out to be useful later.

To simplify the rest of our computations, we re-express the normal mode expansions
of the basic fields [cf. (12)] as [2]
\begin{eqnarray}
A_\mu (x, t) &=& \int dk \; \big[ a_\mu (k) f^*(k, x) + a^\dagger_\mu (k) f(k, x)
\big], \nonumber\\
C (x, t) &=& \int dk \; \big[ c (k) f^*(k, x) + c ^\dagger (k) f(k, x) 
\big], \nonumber\\
\bar C (x, t) &=& \int dk \; \big[ \bar c (k) f^*(k, x)) + \bar c ^\dagger (k) f(k, x)
\big], 
\end{eqnarray}
where the new functions
\begin{eqnarray}
f(k, x) = \frac {e^ {-i k \cdot x}} {\surd {(2 \pi \; 2 k_0)}}, \qquad
f^*(k, x) = \frac {e^ {i k \cdot x}} {\surd {(2 \pi \; 2 k_0)}},
\end{eqnarray}
form an orthonormal set because they satisfy the following conditions
\begin{eqnarray}
\fl \qquad && \int dx \;  f^*(k, x) \; i\, ^\leftrightarrow_{\partial_0} \; f(k', x) (x) 
= \delta (k - k'), \nonumber\\
\fl \qquad && \int dx \; f^*(k, x) \; i\, ^\leftrightarrow_{\partial_0} \; f^*(k', x) = 0, \quad
\int dx \; f(k, x) \; i\, ^\leftrightarrow_{\partial_0} \; f(k', x) = 0.
\end{eqnarray} 
We have taken into account, in the above, the following standard definition 
\begin{eqnarray}
A \; ^\leftrightarrow_{\partial_0} \; B = A (\partial_0 B) - (\partial_0 A) B, 
\end{eqnarray}
for the operator $^\leftrightarrow_{\partial_0}$ between two arbitrary non-zero 
variables A and B. Using the above relations, it is straightforward to check that 
\begin{eqnarray}
\fl \qquad a_\mu (k) = \int dx \; A_\mu (x, t) \; i\, ^\leftrightarrow_{\partial_0}\; f(k, x), \qquad
a^\dagger_\mu (k)  = \int dx \;  f^*(k, x) \; i\, ^\leftrightarrow_{\partial_0}\;
A_\mu (x, t),\nonumber\\
\fl \qquad c(k) = \int dx \; C(x, t) \; i\, ^\leftrightarrow_{\partial_0}\; f(k, x), \;\;\quad \qquad
 \bar c(k) = \int dx \; \bar C(x, t) \; i\, ^\leftrightarrow_{\partial_0}\; f(k, x), \nonumber\\
\fl \qquad c^\dagger (k) = \int dx \; f^*(k, x) \; i\, ^\leftrightarrow_{\partial_0}\; C (x, t), \;\;\;\quad \quad
 \bar c^\dagger (k) = \int dx \; f^*(k, x) \; i\, ^\leftrightarrow_{\partial_0}\; \bar 
C (x, t).
\end{eqnarray}
Thus, we have expressed the creation and annihilation operators in terms of the basic fields
and the orthonormal functions $f(k, x)$ and $f^*(k, x)$.

At this stage, a few important comments are in order. First and foremost, it is straightforward to
check that only the canonical brackets (11) survive in the explicit computations. Second,
there exist six anticommutators from the four fermionic operators $ c(k), \; c^\dagger (k),
\; \bar c(k), \; \bar c^\dagger (k)$. Out of which, four turn out to be zero because of the 
orthonormality relations (59) and due to the fact that $ C^2 = \bar C^2  = 0, \;
\{ C (x, t), \; \dot C (y, t) \} = 0, \;
\{ \bar C (x, t), \; \dot {\bar C} (y, t) \} = 0$. Finally, there exist basic
commutators from the operators $ a_\mu (k)$ and $a_\mu^\dagger (k)$. Out of which, two turn out
to be zero (i.e. $[a_\mu (k), \; a_\nu (k^\prime)] 
= [a^\dagger_\mu (k), \; a^\dagger_\nu (k^\prime)] = 0$). The proof for it is simple
because the commutation relations in (56) can be recast in the form
$[ A_\mu (x, t), \; \dot A_\nu (y, t)] = - i \; \eta_{\mu\nu}\; \delta (x - y)$
due to the fact that (i) $\dot A_0 = (\partial \cdot A) + \partial_i A_i$  and
$\dot A_i = E_i  + \partial_i \; A_0$, and (ii) the spatial derivative of the gauge
field $A_\mu$ commutes with itself (i.e. $[A_\mu (x, t), \; \partial_i A_\nu (y, t) ] = 0$).

It is straightforward to check that the canonical brackets of (53) and (54)
lead to the derivation of {\it exactly} the same brackets as are listed in 
(11). This can be checked directly by exploiting the explicit expressions 
for the creation and annihilation operators quoted in (59) and using the 
canonical brackets listed in (53) and (54). In this 
computation, the concept of normal ordering has {\it not} yet been exploited 
because we have not dealt with any non--trivial physical quantity (e.g.
Hamiltonian, conserved charges, etc.) for our analysis and computation.

The stage is now set to discuss  the canonical quantization in terms of the fermionic 
creation and annihilation operators of the Dirac fields present in the Lagrangian density (2) 
for the 2D QED. As pointed out earlier, we have only the conjugate momentum corresponding 
to the $\psi$ field [i.e. $\Pi_\psi = - i \; \psi^\dagger(x)$]. 
Thus, the analogue of (53) (for the case of Dirac fields) is 
\begin{eqnarray}
\fl \qquad \bigl \{\psi (x, t),\;  \Pi_\psi (y, t) \bigr \} \; = \; i \; \delta(x-y) \Rightarrow  
\bigl \{\psi (x, t), \; \psi^\dagger (y, t) \bigr \} \; = \; - \; \delta(x-y), 
\end{eqnarray}
and rest of the relevant (anti)commutators are zero  amongst the fermionic variables 
($\psi, \psi^\dagger$). Furthermore, the Dirac fermionic fields $\psi$ and $\psi^\dagger$ 
have the zero equal-time (anti)commutators with the rest of the basic fields 
(i.e. $A_\mu, C, \bar C$) and their corresponding conjugate momenta 
that are derived from (2). 
It can be checked that the creation and annihilation operators in the expansions 
of $\psi$ and $\psi^\dagger$ 
[cf. (15), (16)], can be explicitly expressed in terms of these fields
itself [by exploiting the relationships enumerated in (17)]. In their gory
details, these operators can be written as
\begin{eqnarray}
 b^\alpha (k) &=& \int \frac {d k \; e^{- i k \cdot x}}{\surd {(2 \pi \; 2k_0)}} \; 
(u^\alpha )^\dagger (k) \; \psi (x, t),\nonumber\\
 (b^\alpha )^\dagger (k)&=& \int \frac {d k \; e^{+ i k \cdot x}}{\surd {(2 \pi \; 2k_0)}} 
\; \psi^\dagger (x, t) \;  u^\alpha (k), \nonumber\\ 
 d^\alpha (k) &=& \int \frac {d k \; e^{- i k \cdot x}}{\surd {(2 \pi \; 2k_0)}} 
\; \psi^\dagger (x, t)  \; v^\alpha (k), \nonumber\\
 (d^\alpha )^\dagger (k) &=& \int \frac {d k \; e^{+ i k \cdot x}}{\surd {(2 \pi \; 2k_0)}} 
\; (v^\alpha )^\dagger (k) \; \psi (x, t),    
\end{eqnarray}
where we have made use of the useful relations (17) (i.e. $ (u^\alpha )^\dagger (k) \; u^\beta (k) 
= 2 k_0 \; \delta^{\alpha \beta}$, $(v^\alpha )^\dagger (k) \; v^\beta (k) 
= 2 k_0 \; \delta^{\alpha \beta}$). It is now straightforword to check that
\begin{eqnarray}
\{ b^\alpha (k), \; (b^\beta)^\dagger (k^\prime) \} = - \; \delta^{\alpha \beta} 
\; \delta (k - k^\prime), \nonumber\\
\{ d^\alpha (k), \; (d^\beta)^\dagger (k^\prime) \} = - \; \delta^{\alpha \beta} 
\; \delta (k - k^\prime),
\end{eqnarray}
and rest of the (anti)commutators amongst $b^\alpha (k), \;  (b^\beta)^\dagger (k),\; 
d^\alpha (k)$ and $ (d^\beta)^\dagger (k)$ are found to be zero where, in these proofs,
we have to make use of $ \{ \psi (x, t), \; \psi (y, t) \} = 0, \; \{ \psi^\dagger (x, t), 
\; \psi^\dagger (y, t) \} = 0$ and various relations that exist amongst $u^\alpha (k)$, 
$(u^\beta)^\dagger (k)$, $v^\alpha (k)$ and $(v^\beta)^\dagger (k)$ (see, e.g. [13] 
for details). Thus, we obtain the same canonical anticommutators [cf. (28)] as 
derived earlier by exploiting the continuous and nilpotent BRST symmetry transformations [cf. (18)].

Finally, we conclude that the {\it basic} canonical brackets, 
amongst the creation and annihilation operators of the bosonic and fermionic fields of the 
2D QED with Dirac fields, can be derived in a straightforward manner
\begin{enumerate}
\item from the continuous symmetry considerations, and 

\item  by exploiting the definition of momenta from the Lagrangian density.
\end{enumerate}
From both the above methods, the basic brackets [cf. (11), (62)]
turn out to be {\it exactly} the same. Thus, 
the basic brackets (11) and (62) are hidden, in a subtle way, in the
continuous symmetry transformations of our present interacting Hodge theory itself.

\section{Conclusions}
The central result of our present investigation is the derivation of the basic canonical 
brackets by exploiting the continuous symmetry transformations that are present in the
{\it interacting} 2D Abelian 1-form gauge theory with Dirac fields. These brackets exist amongst 
the creation and annihilation operators that appear in the normal mode expansions of the 
basic dynamical  fields of the interacting  theory. In our present endeavor, the key ideas that have been exploited 
are the spin-statistics theorem, normal ordering (in the expressions for the charges) 
and continuous symmetry transformations. The last of the above ideas is a {\it novel} one 
and it differs from the standard method of canonical quantization scheme where the 
{\it classical} (graded) Poisson-brackets (with the {\it mathematical}
 definition of the canonical momenta) 
are promoted to the {\it quantum} (anti)commutators in addition to the
spin-statistics theorem and normal ordering.

It should be noted that, in some of the standard text books (see, e.g. [14]), the 
canonical brackets amongst the creation and annihilation operators have been obtained 
by exploiting the Poincar{\' e} operators (like momenta and angular momenta) which 
are generators of the global {\it spacetime} transformations (i.e. translations plus 
Lorentz rotations). In our case, however, we have exploited only the continuous 
{\it internal} symmetry transformations connected with the BRST formalism. These 
continuous symmetry transformations are needed to prove that the 2D QED with Dirac 
fields is a field-theoretic model for the Hodge theory [8,9] where the above symmetry 
transformations (and corresponding charges) provide the physical realizations of the 
de Rham cohomological operators of differential geometry (see, e.g. [8,9,4]).

One of the most beautiful observations in our present investigation is the emergence 
of one and the same non-vanishing basic canonical brackets [cf. (11), (62)] from 
{\it all} the continuous symmetry transformations  that are present in the theory. 
In fact, even though the continuous transformations [cf. (3), (5), (7), (9)] 
(and their corresponding generators)  
look drastically different, the hidden basic brackets [cf. (11), (62)], that emerge from the 
application of (13), are exactly the same. This key observation ensures that the 
symmetry principles encode in their folds the canonical brackets, too. 
To the best of our knowledge, our method of the derivation of (11) and (62) 
is a novel result in the realm of quantization of the gauge field theories.

We have purposely discussed, separately and independently, the free 2D U(1) gauge theory [3]
and the interacting 2D QED with Dirac fields. This is due to the fact that all the 
conserved charges in the case of the former turn out to be quadratic (bilinear) in 
fields [3]. As a consequence, they can be expressed in terms of the creation and annihilation
operators in a neat and compact form without any exponentials. This is not 
the situation with the 2D QED with Dirac fields. It can be seen that the continuous 
(anti-)BRST and (anti-)co-BRST transformations of the Dirac fields ($\psi(x, t)$
and $\psi^\dagger(x, t)$) are generated by the charges that contain trilinear 
terms which can {\it not} be expressed in terms of the creation and annihilation operators
in a compact and neat fashion. Thus, the derivation of the basic canonical brackets,
from the above trilinear terms, becomes quite involved. However, we have been able 
to get rid of this problem and accomplished our goal in a clear fashion. To demonstrate
that our method of quantization is general in nature, we have applied it to the BRST
quantization of 4D free Abelian 2-form gauge theory [3] which also happens to be
a tractable field theoretic model for the Hodge theory (see, e.g. [6]).

In our present endeavor, we have considered the 2D QED with a background spacetime  manifold
which is flat, Minkowskian and {\it commutative} in nature. We have demonstrated explicitly
the equivalence between the (anti)commutators at the {\it field} level and at the level of 
creation and annihilation {\it operators} that appear in the normal mode expansion of the
dynamical fields. This equivalence breaks down in the case of 2D Minkowskian spacetime
that is taken to be {\it noncommutative}. In fact, in a very interesting piece of 
recent work [15], it has been clearly demonstrated that the results are completely different
when one exploits the coordinate coherent states approach for the discussion of the Unruh effect 
and Hawking radiation by adopting two different kinds of quantization procedures (because it is well-known
that, in context of the noncommutative field theories, the canonical and noncanonical quantization
procedures do exist). However, in our present investigation on the 2D QED, we have {\it not} adopted
the coordinate coherent states approach and our entire discussion is confined to the commutative
spacetime only. Furthermore, our novel approach of quantization procedure is valid only for
the special class of gauge field theoretic models which present tractable 
examples of Hodge theory.

Symmetry principles, as is well--known, have already played decisive roles in the developments 
of modern theoretical physics. We firmly believe that the key aspects of symmetry principles, 
highlighted in our present investigation, can be generalized to the description
of higher $p$-form ($p \geq 2$) gauge fields that appear in the excitations of the 
(super)strings. Thus, our present endeavor should be taken as our modest step towards 
our main goal of studying various aspects of the free 4D Abelian 2-form (see, e.g. [3])
and higher $p$-form ($p \geq 3$) gauge theories within the framework of BRST formalism.
In this context, it is gratifying to point out that we have already applied the
idea of our present work to the free 4D Abelian 2-form gauge theory and derived 
the correct basic canonical brackets amongst the creations and annihilation operators
by exploiting the continuous symmetries of the theory [3]. We hope
to pursue, some of the above mentioned issues (especially related with
the higher $p$-form ($p \geq 3$) gauge theories), in the future, too [16].

\ack 
SG and RK would like to gratefully acknowledge the financial
support from CSIR and UGC, New Delhi, Government of India, respectively.

\appendix
\section{}

We establish here  the consistency between the anticommutators of
various types amongst the creation and annihilation operators of 
the Dirac fields and the ones that exist between the field variables
themselves (i.e. $ \{ \psi (x, t), \; \psi (y, t) \} = 0, \;
\{ \psi^\dagger  (x, t), \; \psi^\dagger (y, t) \} = 0, \;
\{ \psi (x, t), \; \psi^\dagger (y, t) \} = - \;\delta (x - y)$). 
To this end in mind, it can be seen that the l.h.s. of the last  anticommutator 
can be expressed, using expansions (15) and (16), as
\begin{eqnarray}
\fl \qquad && \{ \psi (x, t),\; \psi^\dagger (y, t) \} =  
{\displaystyle \frac{1}{2 \pi} \;\int   \frac { d k \;d k^\prime}{ \surd (2 k_0 \; 2 k_0^\prime)} 
\sum_\alpha\sum_\beta} \nonumber\\
\fl \qquad && \Bigl [ \bigl \{ b^\alpha (k),\; (b^\beta)^\dagger (k^\prime) \bigr \}\;
u^\alpha (k) \; (u^\beta)^\dagger (k^\prime) \; 
exp\; [{i (k_0 -  k_0^\prime) \; t - i (k x - k^\prime y)}] \nonumber\\
\fl \qquad && + \bigl \{ (d^\alpha)^\dagger (k),\;  d^\beta (k^\prime) \bigr \} \; 
v^\alpha (k) \; (v^\beta)^\dagger (k^\prime) \;
exp\; [{-i (k_0 -  k_0^\prime)\;  t + i (k x - k^\prime y)}] \nonumber\\
\fl \qquad && + \bigl \{ (d^\alpha)^\dagger (k),\;  (b^\beta)^\dagger (k^\prime) \bigr \}\;  v^\alpha (k) 
\; (u^\beta)^\dagger (k^\prime) \; 
exp\; [{- i (k_0 +  k_0^\prime) \; t + i (k x + k^\prime y)}]  \nonumber\\
\fl \qquad && + \bigl \{ b^\alpha (k),\;  d^\beta (k^\prime) \bigr \}\;  u^\alpha (k) 
\; (v^\beta)^\dagger (k^\prime) \;exp \;[{i (k_0 +  k_0^\prime) \; t 
- i (k x + k^\prime y)}] \Bigr ],
\end{eqnarray} 
where we have {\it not} taken $ t = 0 $ for the sake of generality and transparency.
It is obvious, from this canonical bracket, that the r.h.s. of the above equation is the 
Dirac delta-function [cf. (29)]. A comparison with the definition of the delta-function 
implies that the following anticommutators are true:
\begin{eqnarray}
\{ b^\alpha (k), \; d^\beta (k^\prime) \} = 0, \qquad 
\{ (b^\alpha)^\dagger (k),\;  (d^\beta)^\dagger (k^\prime) \} = 0.
\end{eqnarray}
The above conclusion is drawn because of the exponentials that are present in the 
Dirac delta-function (29) (i.e. r.h.s) and (A.1) (i.e. l.h.s.). It can be easily seen that the 
exponentials in third and fourth terms of (A.1) can not produce $\delta (x - y)$ 
for $k$ and $k^\prime$ being positive definite.

Furthermore, if we assume the other remaining brackets to be (62), we obtain the 
following expression for (A.1), namely;
\begin{eqnarray}
\fl \qquad &-& {\displaystyle \frac{1}{2 \pi} \;\int   \frac { d k \;d k^\prime}{ \surd (2 k_0 \; 
2 k_0^\prime)} \sum_{\alpha}}\; \delta (k - k^\prime) \;
\Bigl (u^\alpha (k) \; (u^\alpha)^\dagger (k^\prime)\;
exp \;[- \; i \; k \; (x - y) ]  \nonumber\\
\fl \qquad &+&  v^\alpha (k) \; (v^\alpha)^\dagger (k^\prime)\;
exp \;[+ \; i \; k \; (x - y) ]   \Bigr ),  
\end{eqnarray}
where we have taken into account the fact that $k_0 = k^\prime_0$ because of the presence 
of the $\delta (k - k^\prime)$  [implying $e^{i (k_0 - k^\prime_0) \; t} \; 
\delta (k - k^\prime) = \delta (k - k^\prime)$].
Inserting $(\gamma^0)^2 = I$ appropriately and using (17), we obtain the following
\begin{eqnarray}
&&- {\displaystyle \frac{1}{2\pi} \;\int \frac{dk} {2 k_0}}\;
\Bigl [(k_0 \gamma_0 - k \gamma_1 + m)\; \gamma_0 \; exp\; [- \; i \; k \; (x - y)] \nonumber\\
&& + \; (k_0 \gamma_0 - k \gamma_1 - m)\; \gamma_0 \; exp\; [+ \; i \; k \; (x - y)] .
\end{eqnarray}
Changing $ k \to - k$ in the second term, we obtain the following final expression
\begin{eqnarray}
- {\displaystyle \int  \frac{d k} {(2 \pi) (2  k_0)}} \; (2 k_0) \; 
exp \; [- \; i \; k \; (x - y)],
\end{eqnarray}  
which is nothing other than the Dirac delta-function.

Similarly, it can be checked that there is an absolute consistency between the canonical
anticommutators that emerge from the Lagrangian density (2):
\begin{eqnarray}
\bigl \{ \psi (x, t),\;  \psi (y, t) \bigr \} \; = \; \bigl \{ \psi^\dagger (x, t),\; 
 \psi^\dagger (y, t) \bigr \} \; = \; 0,
\end{eqnarray} 
and the following anticommutators amongst the creation and annihilation operators
that ensue due to the mode expansions of the above fields, namely;
\begin{eqnarray}
&& \bigl \{ b^\alpha (k),\;  b^\beta (k^\prime) \bigr \} \; = \;  0, \qquad \qquad \; 
\bigl \{ d^\alpha (k),\;  d^\beta (k^\prime) \bigr \} \; = \; 0, \nonumber\\ 
&& \bigl \{ b^\alpha (k),\;  (d^\beta)^\dagger (k^\prime) \bigr \} \; = \; 0,
\qquad \quad \;  \bigl \{ (b^\alpha)^\dagger (k), \; (b^\beta)^\dagger (k^\prime) \bigr \}
 \; = \; 0, \nonumber\\
&& \bigl \{ (d^\alpha )^\dagger (k),\;  (d^\beta)^\dagger (k^\prime) \bigr \} \; = \; 0, \qquad
\bigl \{ (b^\alpha)^\dagger (k), \; d^\beta (k^\prime) \bigr \} \; = \; 0.
\end{eqnarray}
Thus, we have demonstrated an absolute consistency between the canonical brackets consisting 
of $\psi (x, t)$ and $\psi^\dagger (x, t)$ and the corresponding canonical brackets 
(anticommutators) existing amongst the creation and annihilation operators that
appear in the mode expansions of the above fields. In other words, we have
established clearly the equivalence between the canonical brackets at the {\it field} level and
the corresponding brackets at the level of the fermionic
{\it creation} and {\it annihilation} operators which appear in the normal mode expansions of the 
fermionic Dirac fields.

\end{document}